\documentclass[preprintnumbers,amsmath,amssymb]{revtex4}
\usepackage{graphicx}
\usepackage{dcolumn}
\usepackage{bm}
\def\eqnn#1{(\ref{eq:#1})}

\def\jsp{{\sl  J. Stat. Phys.}}
\def\figno#1{Fig.~\ref{fig:#1}}
\def\cum#1{\langle\langle#1\rangle\rangle}
\def\vev#1{\langle#1\rangle}
\def\tkick{t_{kick}}
\begin{document}
\title{Violations of local equilibrium and stochastic thermostats} 
\author{
  Kenichiro Aoki\footnote{E--mail:~{\tt ken@phys-h.keio.ac.jp}.
      Supported in part by the Grant--in--Aid for Scientific Research
from the Ministry of Education, Culture, Sports, Science and
Technology of Japan.}}
\affiliation{Dept. of Physics, Keio University, {\it
    4---1---1} Hiyoshi, Kouhoku--ku, Yokohama 223--8521, Japan}
  \date{\today }

\begin{abstract}
    We quantitatively investigate the violations of local equilibrium
    in the $\phi^4$ theory under thermal gradients, using stochastic
    thermostats. We find that the deviations from local equilibrium
    can be quite well described by a behavior $\sim(\nabla T)^2$. The
    dependence of the quantities on the thermostat type is analyzed
    and its physical implications are discussed.
\end{abstract}

\vspace{3mm}

\maketitle

The physics of non-equilibrium is ubiquitous and is an essential
ingredient in a wide spectrum of issues in physics, from the early
universe, transport in solid state physics to heavy ion
collisions.
Because of this importance, the field has been studied for a long
time, yet some basic problems remain\cite{neqRevs}. A deep and
interesting issue is what the microscopic properties of the systems in
non--equilibrium are.
In particular, a crucial question is whether  local 
equilibrium is broken when global equilibrium is broken, and if it is,
how the breaking occurs.  
Non--equilibrium situations occur in various situations and there is
no general theory on how systems should behave away from equilibrium.
In steady state systems, the validity of local equilibrium has been
addressed from various approaches, though few have studied the
deviations from it
quantitatively\cite{fluct,eos,local-eq,onsager,hk,dhar,ak-le}.
The violations of local equilibrium were studied quantitatively in the
FPU (Fermi--Pasta--Ulam) model and the $\phi^4$ theory under thermal
gradients through the
behavior of momentum cumulants, making use of numerical
simulations\cite{ak-le}. There, the violations of local equilibrium
were seen and their dependence on the non-equilibrium nature of the
system could be well described by $(\nabla T)^2$, where $T$ is the
local temperature.  Furthermore, it was found that the linear response
theory also breaks down by about the same amount, relatively.  These
non--equilibrium states were numerically constructed using
deterministic thermostats in a standard manner.
However, it has not yet been established whether these deviations also
occur if completely different type of thermostats are used, such as
stochastic ones.  Since the non--equilibrium situation itself is
created by the action of the thermostats, whether the non--equilibrium
effects are artifacts of using a particular type of thermostat needs
to be clarified.

The possible dependence of physics properties on thermostats is not a
purely theoretical concept; the physical quantities can depend on the
thermostats in a non-trivial way, in real physical systems.  A
thermostat thermalizes the degrees of freedom attached to it at a
prescribed temperature. However, what happens away from the
thermostatted regions is not unique and is a dynamical property of the
system.
For instance, it has been  known for some time that the internal
gradient (be it shear or heat flow), depends on the boundary
conditions in a non-trivial way through the boundary jumps or
slips.
For thermostats, these slips depend not only on the temperature but
also on the type of thermostat used, which is quite understandable if
one considers real systems.
The issue  involves a deeper question  --- which physical
quantities are universal or intrinsic to the system, independently of
the boundary conditions. 
In particular, for systems with bulk behavior, physical quantities
that are universal in this sense should be independent of the size of
the system and the thermostat used. 
The boundary conditions determine the flow
and gradients inside the system, away from the boundaries, but whether
their influence on physical quantities is solely through these
quantities is unclear.
Determining which quantities are universal in the sense that they are
independent of the thermostats  and
the boundary conditions is crucial if one wants to develop general
theories regarding the behavior of systems in non--equilibrium, such
as in the approach of so--called non--equilibrium
thermodynamics\cite{neqThermo}.

In this work, we quantitatively investigate the steady state
non--equilibrium behavior of the $\phi^4$ theory in one spatial
dimension using stochastic thermostats. (For a review of one
dimensional systems
see, for instance,
\cite{1dreview}.) We find that the behavior is similar to that of
systems using deterministic thermostats, but differs in a subtle way.
The Hamiltonian of the massless $\phi^4$ theory in one dimension,
which  we study, is 
\begin{equation}
  \label{eq:ham}
  H=\sum_{j=1}^L\left[{\pi_{j}^2\over2} +
    {\left(\nabla \phi_{j}\right)^2\over2} 
    + {\phi_{j}^4\over4}\right]
\end{equation}
with $L$ being the size of the system and the discrete spatial
gradient defined as $\nabla\phi_j\equiv \phi_{j+1}-\phi_j$.  
$\phi^4$ theory is a typical model in which bulk behavior can be seen
in the $(1+1)$ dimensional model. The model is a prototypical one in
field theory so that it allows us to understand the problem in a
broader context and it is applicable to various problems in
solid state and particle physics.
We adopt the so called ``ideal gas
temperature'', $\vev{\pi_j^2}=T_j$ as the definition of temperature.  This
widely used definition is convenient for systems we study, since it is
simple and local. 
We note that it is quite sensible to consider local
quantities in the system as we are considering physical quantities 
averaged over a large number of ensembles.  

In this study, we analyze the $\phi^4$ theory under thermal gradients,
where thermostats are applied to one site each at both ends of the
system, with their temperatures $T_{1,2}^0$.  Inside the system, the
equations of motion are solely those derived from the Hamiltonian
\eqnn{ham}. We integrate the equations of motion numerically computing
the values of relevant observables at every step and averaging them
over time at the end in the standard manner. We used time steps of
$0.005\sim0.05$ over $10^9\sim10^{10}$ steps and have checked that the
results do not depend on the time step size.

We apply the stochastic
thermostats following the ideas in \cite{Jackson68}.  Concretely, at
every $\tkick$ time interval, we replace the momenta at the
boundaries, $\pi_1,\pi_L$, with values selected randomly from the
Maxwell distribution for the thermostat temperatures, $T_{1,2}^0$.
Physically, these thermostats can be thought of as particles at the
boundaries of the chain striking the walls at given temperatures,
though the situation is more general since $\phi_j$ needs not have any
relation to a spatial direction. For instance, in the lattice $\phi^4$
field theory, $\phi_j$ is an internal degree of freedom, not a spatial
one.  $\tkick$ affects how well the thermostats work and has to be
controlled so that thermalization is achieved at the boundaries.
Given the boundary temperatures, one still needs to specify the type
of thermostat to apply and this is not just a theoretical choice;
given the boundary temperatures, the choice of the type of thermostat
used affect the physical quantities in real systems also.  If the
system displays bulk behavior, quantities such as thermal conductivity
should have no thermostat dependence.

\begin{figure}[hptb]
  \centering
   \includegraphics[width=9cm,clip=true]{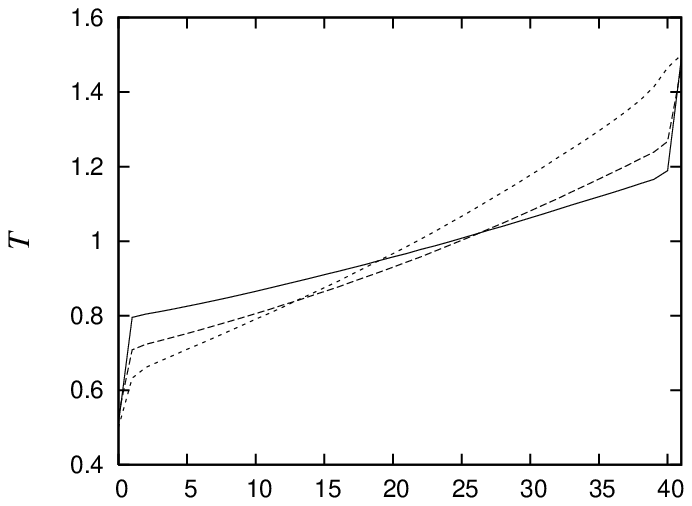}\\
   \includegraphics[width=9cm,clip=true]{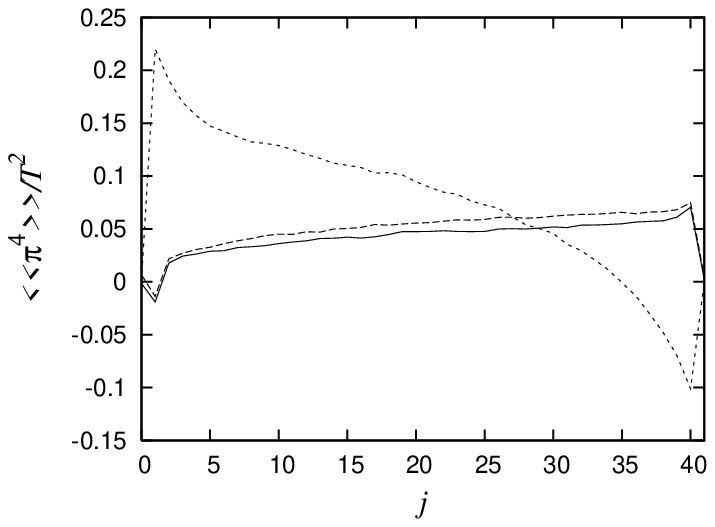}
   \caption{Temperature profile (top) and $\cum{\pi^4}/T^2$ cumulant
     profile (bottom) for boundary temperatures
     $(T_1^0,T_2^0)=(0.5,1.5)$. Thermostats used were stochastic with
     $\tkick=0.5$ (solid), $\tkick=1$ (dashes) and deterministic
     (small dashes). }
  \label{fig:tprofs}
\end{figure}
In \figno{tprofs}, we show some spatial profiles of the temperature
and $\pi^4$ cumulants, $\cum{\pi_k^4} = \langle \pi_k^4\rangle -
3\langle \pi_k^2\rangle^2$, with the boundary temperatures controlled
by the stochastic thermostats.  The profiles include temperature jumps
at the boundaries and thermal profiles inside which can be curved.
These features are quite generic. Given a Maxwellian distribution, all
the cumulants vanish, except for $\cum{\pi^2}$. When the distribution
is not Maxwellian, there is no unique definition of $T$ and in this
sense, local equilibrium does not hold\cite{nonEqT}.  
The cumulant profiles show that the boundaries are indeed thermalized
but local equilibrium is broken inside the system.  The behavior of
the cumulants are quite different but the corresponding temperature
profiles are also quite different.  We analyze these properties in
more detail below.
Considering cumulants of $\pi$ in non-equilibrium is reminiscent of
Grad's moment expansion in kinetic theory of some time
ago\cite{neqThermo}. In that approach, one tries to obtain the
distribution function $f(\pi_j,\phi_k)$ in non-equilibrium by
constraining the coefficients the expansion in $\pi$ and $\phi$ using
macroscopic observables.

To study the dynamics of the system, let us first measure the thermal
conductivity, $\kappa$,  directly. We perform this task by choosing
temperature boundary conditions  $(T-\delta T,T+\delta T)$ with
varying $\delta T$ around the same central temperature. In this
manner, we can verify directly  that Fourier's law, $J=-\kappa \nabla
T$ holds in the  non--equilibrium steady state system and  obtain the
thermal conductivity.  Here, $J=-\pi\nabla \phi_i$ is the heat flow in
the system, which is constant inside, since the flow is one dimensional.

\begin{figure}[h]
  \centering
  \includegraphics[width=9cm,clip=true]{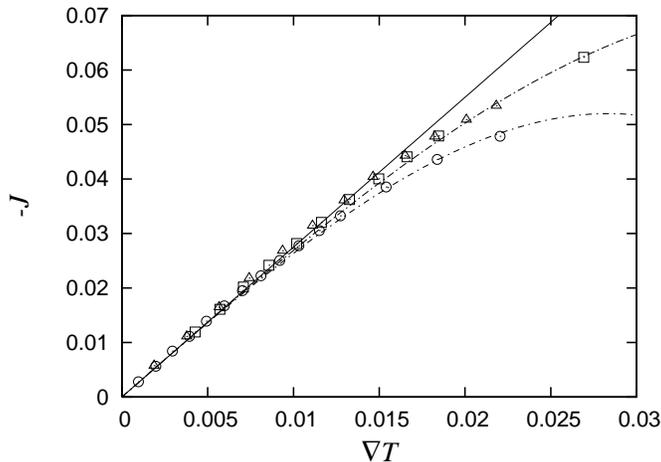}
  \caption{$J$ vs. $\nabla T$. The relation is linear when the system
    is not too far from equilibrium, showing that Fourier's law holds.
    Thermostats used were stochastic with $\tkick=1$ ($\Box$),
    $\tkick=0.5$ ($\bigcirc$) and deterministic ($\bigtriangleup$).
    Straight line represents bulk conductivity and we see that the the
    data all agree, independently of the thermostat type in the linear
    regime, as expected.
    The curved lines represent \eqnn{lr-breaking} (see text).
  }
  \label{fig:fourier}
\end{figure}
In \figno{fourier}, the relation between $J$ and $\nabla T$ is shown
for $L=42$ systems with central temperature $T=1$. 
We have plotted the cases using stochastic
thermostats with $\tkick=0.5,1$ and for comparison, deterministic
thermostats which generalize the Nos\'e--Hoover
thermostats\cite{nose-hoover}, as detailed in \cite{ak-long}.  We see
that Fourier's law is obeyed for systems not too far from equilibrium.
The straight line is Fourier's law in the bulk behavior regime with
$\kappa=2.75$ obtained previously using deterministic
thermostats\cite{ak-long}.  It can be seen that the agreement with the
results using both types of stochastic thermostats is excellent, as we
expect.  At this temperature, bulk behavior in thermal conductivity
holds quite well at $L=42$. The thermal conductivity is
independent of both system size and of thermostat type and is
universal.

For systems far from equilibrium, we see that the linear response
theory starts to break down\cite{ak-le}.  The manner in which it breaks
down is similar to the case with deterministic thermostats and can be
reasonably well depicted by the relation,
\begin{equation}
  \label{eq:lr-breaking}
  \delta_{\scriptsize LR}=  {J-J_{LR} \over J_{LR}} = 
  C_{LR}\left(\frac{\nabla T}{T}\right)^2
\end{equation}
This is shown in \figno{fourier}, with $C_{LR}=-414$ (dashes), $-215$
(dot--dashes) for $t_{kick}=0.5,1$ cases respectively.
It should be emphasized that to analyze the deviations from linear
response theory, we also need to consider the possible breaking of
local equilibrium.  Unless local equilibrium is preserved, the meaning
of temperature is not unique so that the significance of linear
response breaking implicitly depends on the definition of the
temperature.

To this end, we quantitatively investigate the deviations from local
equilibrium and we do so through the momentum cumulants,
$\cum{\pi^4}/T^2$. 
In the $\phi^4$ theory and the FPU model thermostatted by
deterministic thermostats, it was found that the local equilibrium
breaking can be described by a simple behavior \cite{ak-le},
\begin{equation}
  \label{eq:le-breaking}
  \delta_{\scriptsize LE}=\frac{ \cum{\pi^4}}{3T^2}= 
  C_{LE} \left(\frac{\nabla T}{T}\right)^2
\end{equation}
\begin{figure}[hptb]
  \centering
   \includegraphics[width=9cm,clip=true]{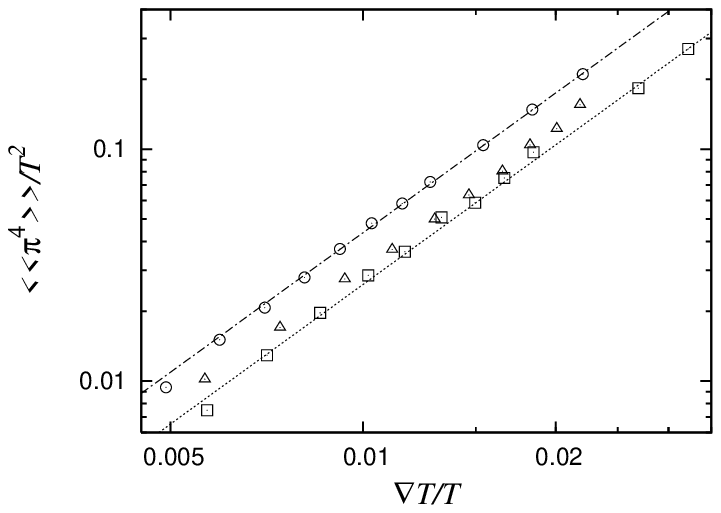}
   \caption{Thermal and $\cum{\pi^4}/T^2$ cumulant for $T=1$ vs.
     $\nabla T/T$.  Thermostats used were stochastic with $\tkick=1$
     ($\Box$), $\tkick=0.5$ ($\bigcirc$) and deterministic
     ($\bigtriangleup$). } 
  \label{fig:cum4}
\end{figure}
In \figno{cum4}, we plot the behavior of $\cum{\pi^4}/T^2$ at $T=1$
against $\nabla T/T$, for $L=42$. 
To do so, we averaged the cumulants for small
number of lattice sites having local temperature close to
$T=1$ inside the system.  By ``inside'' here, we refer to  lattice
sites away by more than the mean free path from the boundaries. The
mean free path and other properties of this system can be found in
\cite{ak-long}.
In the plot, systems using stochastic boundary thermostats with
$\tkick=0.5,1$  are included.  Fits to the behavior
of \eqnn{le-breaking}, using coefficients $C_{LE}= 146, 87$,
respectively, are also shown.
For comparison, we also included the behavior of the case with
deterministic thermostats similar in type to those in \cite{ak-long}.
In the plot, each point represents an averaged cumulant from a
configuration with a particular set of temperature boundary
conditions.  We see that the behavior is quite well described by the
formula \eqnn{le-breaking} in all cases.  For both $C_{LE},C_{LR}$,
$t_{kick}=0.5$ case is larger than $t_{kick}=0.5$ case by a factor of
roughly two, but its precise reason is unclear.

So we see that the breaking of local equilibrium also occurs when
using stochastic thermostats, which are quite different from
deterministic ones, and that the behavior of the breaking is
semi--quantitatively the same.
However, we also see that the behavior is {\it non-universal} in that
the coefficient $C_{LE}$ depends on the thermostat type. While the
behavior of local equilibrium breaking itself is generic and simple,
the coefficient, though of the same order, is not universal. 
When using deterministic thermostats, the coefficient $C_{LE}$ is
dependent on $L$ even though the theory has bulk behavior\cite{ak-le},
so that dependence on thermostat type is perhaps not surprising.
On the technical side, we have found it significantly more demanding to
obtain reliable numerical simulation data from stochastic simulations
than from deterministic ones.

Let us discuss the significance of the non--universality in the
coefficient $C_{LE}$.
For contrast, let us first consider the thermal
conductivity, which {\it is} universal or intrinsic to the theory.
Thermal conductivity can be obtained from the temperature profile and
the current $J$. But both these quantities are strongly boundary condition
dependent; they depend on the system size and the boundary
temperatures. Furthermore, they depend also on the thermostat type
since the boundary temperature jumps depend on it and hence the
temperature profile. Thermal conductivity is boundary condition
independent only in the sense that when we inspect the relation
between $J$ and $T$, Fourier's Law $J=-\kappa \nabla T$ holds 
and $\kappa$ is independent of $L$, boundary temperatures and
thermostat type.  Had we not tried to use $\nabla T$ or looked at the
appropriate relation, we could have concluded that the thermal
transport properties are non--universal.

Now consider the cumulant: This is a property of the local
distribution function. Let us define the full distribution function,
$f(\pi_1,\pi_2,\ldots\pi_L;\phi_1,\phi_2,\ldots\phi_L)$ which is a
function of all physical degrees of freedom. The local distribution
function for $\pi$ is obtained by integrating out all other degrees of
freedom.
\begin{equation}
  \label{eq:flocal}
  f_j(\pi_j) = \int \prod_{k\not=j} d\pi_k\prod_{k}d\phi_k\,
  f(\pi_1,\pi_2,\ldots\pi_L;\phi_1,\phi_2,\ldots\phi_L)
\end{equation}
Consider a site labelled by $j$, away from the boundaries, so that
there are no direct boundary effects.  The question whether this local
distribution is universal is the question whether this distribution
depends on the boundary conditions only through physical variables
such as $T$ and $\nabla T$. For fixed $T$, we have only the $\nabla T$
dependence whose behavior is similar but manifestly thermostat
dependent.
We can consider various possibilities regarding the significance of
the results: 
\begin{enumerate}
\item The behavior {\it is} universal if we consider relations
  involving variables other than $T$ and $\nabla T$.  However, we note
  that quantities such as $\nabla^2 T$ can be reexpressed in terms of
  $\nabla T$, using perturbation theory in linear response
  theory\cite{ak-le}.
\item The behavior is {\it non-}universal; there is no way to
  remove this thermostat dependence and the results cannot be
  rephrased in a thermostat independent manner.
\item The behavior is a small size effect and large systems display
  different behavior that is universal. 
  However, we
  have studied system sizes up to few hundred sites using
  deterministic thermostats and have found $L$ dependent
  behavior\cite{ak-le} and also to some extent using stochastic
  thermostats.
\end{enumerate}
We add here that the maximum Lyapunov exponent for the $\phi^4$
theory is another physical property of the distribution function that
also depends explicitly on the thermostat type \cite{ak-dloss}.

In this work, we analyzed the microscopic properties of $\phi^4$
theory under thermal gradients, in steady state. The thermal gradients
were created by stochastic thermostats placed at the ends of the
system. The semi-quantitative non-equilibrium behavior of the system
is identical to the behavior obtained using deterministic thermostats.
Namely, not only global equilibrium but local equilibrium is broken by
thermal gradients, which can be measured through the momentum
cumulants. Also Fourier's law does not hold far from equilibrium.  The
dependence of these deviations on the non--equilibrium nature can be
described by $\sim (\nabla T/T)^2$ and their coefficients are of the
same order independently of the thermostat type. 
So these properties are thermostat independent for $\phi^4$
theory. The exact coefficients of these deviations, however, have some
thermostat dependence.

It would be interesting to find out how other theories behave
microscopically in non-equilibrium and how the behavior depends on the
thermostats, or perhaps more importantly, which parts do not. We
believe that these questions are of import in understanding the nature
of the non-equilibrium state from fundamental principles.  We have
also studied the FPU model and have found similar behavior, though
numerical precision is harder to achieve. 


\begin{thebibliography}{99}
  \bibitem{neqRevs} See, {\it e.g.}, P. Gaspard, {\sl Chaos,
      Scattering and Statistical Mechanics}, (Cambridge Univ. Press,
    Cambridge, New York, 1998); J. R. Dorfman, {\sl An Introduction to
      Chaos in Nonequilibrium Statistical Mechanics} (Cambridge Univ.
    Press, Cambridge, 1999).
\bibitem{fluct}   A.~Tenenbaum, G. Ciccotti, R. Gallico, {\sl
  Phys. Rev. }{\bf A25} (1982) 2778.
\bibitem{eos} G. Ciccotti, A. Tenenbaum, {\sl J. Stat. Phys.}
  {\bf 23} (1980) 767; C. Trozzi, G. Ciccotti, {\sl Phys. Rev.}
  {\bf A29} (1984) 916.
\bibitem{local-eq}
  W. Loose, G. Ciccotti, {\sl Phys. Rev. }{\bf A45} (1992) 3859;
  M. Mareschal, E.~Kestemont, F.~Baras, E.~Clementi, G.~Nicolis,
  {\sl Phys. Rev. }{\bf A35} (1987) 3883;
  A.~Tenenbaum, {\sl Phys. Rev. }{\bf A28} (1983) 3132.
\bibitem{onsager} R. M. Velasco, L.S. Garcia-Colin, {\sl
  J. Noneq. Thermo.} {\bf 18} (1993) 157.
\bibitem{hk} B. Hafskjold, S.K. Ratkje, \jsp{\bf 78} (1995) 463
\bibitem{dhar}    A. Dhar, D. Dhar, \prl{\bf 82} (1999) 480
\bibitem{ak-le}
  K. Aoki, D. Kusnezov, {\sl Phys. Lett. }{\bf A309} (2003) 377
  \bibitem{neqThermo} See, {\it e.g.}, S.R.~de Groot, P. Mazur, {\sl
      ``Non-equilibrium Thermodynamics''}, North-Holland, Amsterdam
    (1962); J. Keizer, ``Statistical Thermodynamics of Nonequilibrium
    Processes'' (Springer, New York, 1987); D. Jou, G. Lebon, J.
    Casas-Vazques, {\sl ``Extended Irreversible Thermodynamics''}
    (Springer, Berlin, 1996).
\bibitem{1dreview}   S. Lepri, R. Livi, A. Politi, {\sl Phys. Rep. }
  377 (2003) 1
\bibitem{Jackson68} E.A. Jackson, J.R. Pasta, J.F. Waters,
  J. Comp. Phys. 2 (1968) 207
\bibitem{nonEqT} J. Casas-Vazquez, D.Jou,  {\sl Rep.  Prog. Phys.}, {\bf 66}
  (2003)  1937 and references therein.
\bibitem{nose-hoover}  S.~Nos\'{e}, J.~Chem.~Phys. {\bf 81}, 511
  (1984); {\sl Mol.~Phys.} {\bf 52} (1984)
    255 ; W.~G.~Hoover, {\sl Phys.~Rev. }{\bf A 31} (1985) 1695
\bibitem {ak-long}
  K. Aoki, D. Kusnezov,     {\sl Ann. Phys. } {\bf 295} (2002) 50;
  {\sl Phys. Lett. }{\bf B477} (2000) 348
\bibitem{ak-dloss}  K. Aoki, D. Kusnezov, \prd{\bf E68} (2003) 056204
\end{thebibliography}
\end{document}